\documentclass[reprint, amsmath, amssymb, aps, prx, superscriptaddress, floatfix, nofootinbib, citeautoscript, longbibliography]{revtex4-2}
\usepackage[utf8]{inputenc}
\usepackage[T1]{fontenc}
\usepackage[parfill]{parskip}
\usepackage{siunitx}
\usepackage{blindtext}
\usepackage[colorlinks, linkcolor=blue, citecolor=blue, urlcolor=blue, breaklinks=true]{hyperref}
\usepackage{graphicx}
\usepackage{dcolumn}
\usepackage{bm}
\usepackage{color}
\usepackage{units}
\usepackage{dsfont}
\usepackage{hyperref}
\usepackage{xcolor}

\begin{document}

\title{Spectral Vector Beams for High-Speed Spectroscopic Measurements}

\author{Lea Kopf}
\email{lea.kopf@tuni.fi}
\affiliation{Tampere University, Photonics Laboratory, Physics Unit, Tampere, FI-33014, Finland}

\author{Juan R. Deop Ruano}
\affiliation{Tampere University, Photonics Laboratory, Physics Unit, Tampere, FI-33014, Finland}

\author{Markus Hiekkamäki}
\affiliation{Tampere University, Photonics Laboratory, Physics Unit, Tampere, FI-33014, Finland}

\author{Timo Stolt}
\affiliation{Tampere University, Photonics Laboratory, Physics Unit, Tampere, FI-33014, Finland}

\author{Mikko J. Huttunen}
\affiliation{Tampere University, Photonics Laboratory, Physics Unit, Tampere, FI-33014, Finland}

\author{Frédéric Bouchard}
\affiliation{National Research Council of Canada, 100 Sussex Drive, Ottawa, Ontario K1A 0R6, Canada}

\author{Robert Fickler}
\email{robert.fickler@tuni.fi}
\affiliation{Tampere University, Photonics Laboratory, Physics Unit, Tampere, FI-33014, Finland}

\begin{abstract} 
Structured light harnessing multiple degrees of freedom has become a powerful approach to use complex states of light in fundamental studies and applications. 
Here, we investigate the light field of an ultrafast laser beam with a wavelength-depended polarization state, a beam we term spectral vector beam. 
We demonstrate a simple technique to generate and tune such structured beams and demonstrate their spectroscopic capabilities. 
By only measuring the polarization state using fast photodetectors, it is possible to track pulse-to-pulse changes in the frequency spectrum caused by, e.g. narrowband transmission or absorption. 
In our experiments, we reach read-out rates of around 6\,MHz, which is limited by our technical ability to modulate the spectrum and can in principle reach GHz read-out rates. 
In simulations we extend the spectral range to more than 1000\,nm by using a supercontinuum light source, thereby paving the way to various applications requiring high-speed spectroscopic measurements.
\end{abstract}

\maketitle

\section{Introduction} 

Spectroscopic techniques are amongst the most important experimental optical methods with applications ranging from physics to chemistry, material science, and biology \cite{gauglitz2014handbook}. 
In optical spectroscopy, the change of the frequency spectrum of light due to an interaction with matter allows to draw conclusions about the properties of a sample.
In many cases, the progress in spectroscopic methods was initiated by the emergence of novel states of light interacting with the object under investigation \cite{tkachenko2006optical}.

Over the last decades, various ways have been conceived to structure light in complex manners, which have involved different degrees of freedom, such as the temporal \cite{weiner2011ultrafast} and spatial domain \cite{rubinsztein2016roadmap}.
By correlating different properties, the complexity has further been increased, e.g.~by combining time and polarization \cite{Walecki1997Characterization,Villeneuve2000Forced}, or time and space \cite{koehl1998automated,shaltout2019spatiotemporal}.
In both fields, the control and understanding over the spatio-temporal shape of light have enabled novel fundamental studies and a myriad of applications \cite{weiner2011ultrafast, forbes2016creation}.
Linking time and polarization brought up new technological opportunities which have been used, for example, in molecular optics \cite{Villeneuve2000Forced} and polarimetric measurements \cite{Pinnegar2006Polarization,Fade2012Depolarization}, while frequency--polarization linked beams have been utilized to realize wavemeters \cite{Sano1980Simple,Dimmick1997Simple}.

Amongst these ways to increase the complexity of the light field, one particularly interesting example is the so-called spatial vector beam.
It features a varying transverse spatial amplitude and a varying polarization vector across the beam extent \cite{dennis2009singular,rosales2018review}.
Not only fundamental studies \cite{bauer2015observation,larocque2018reconstructing} and various applications in classical \cite{dennis2009singular,rosales2018review} and quantum optics \cite{rubinsztein2016roadmap} of such beams have been of interest, but also analogies between the two domains \cite{spreeuw1998classical,aiello2015quantum,forbes2019classically}.
A recent study shows that strong correlations between the transverse position and polarization in spatial vector beams can be used for high-speed kinematic sensing \cite{Berg-Johansen2015Classically}. 
There, an object that is moving through the beam in the transverse direction changes the overall polarization of the beam such that simple polarization measurements can be used to track the motion of the object with high-speed. 

\begin{figure}[h] 
  \centering
  \includegraphics[width=0.95\linewidth]{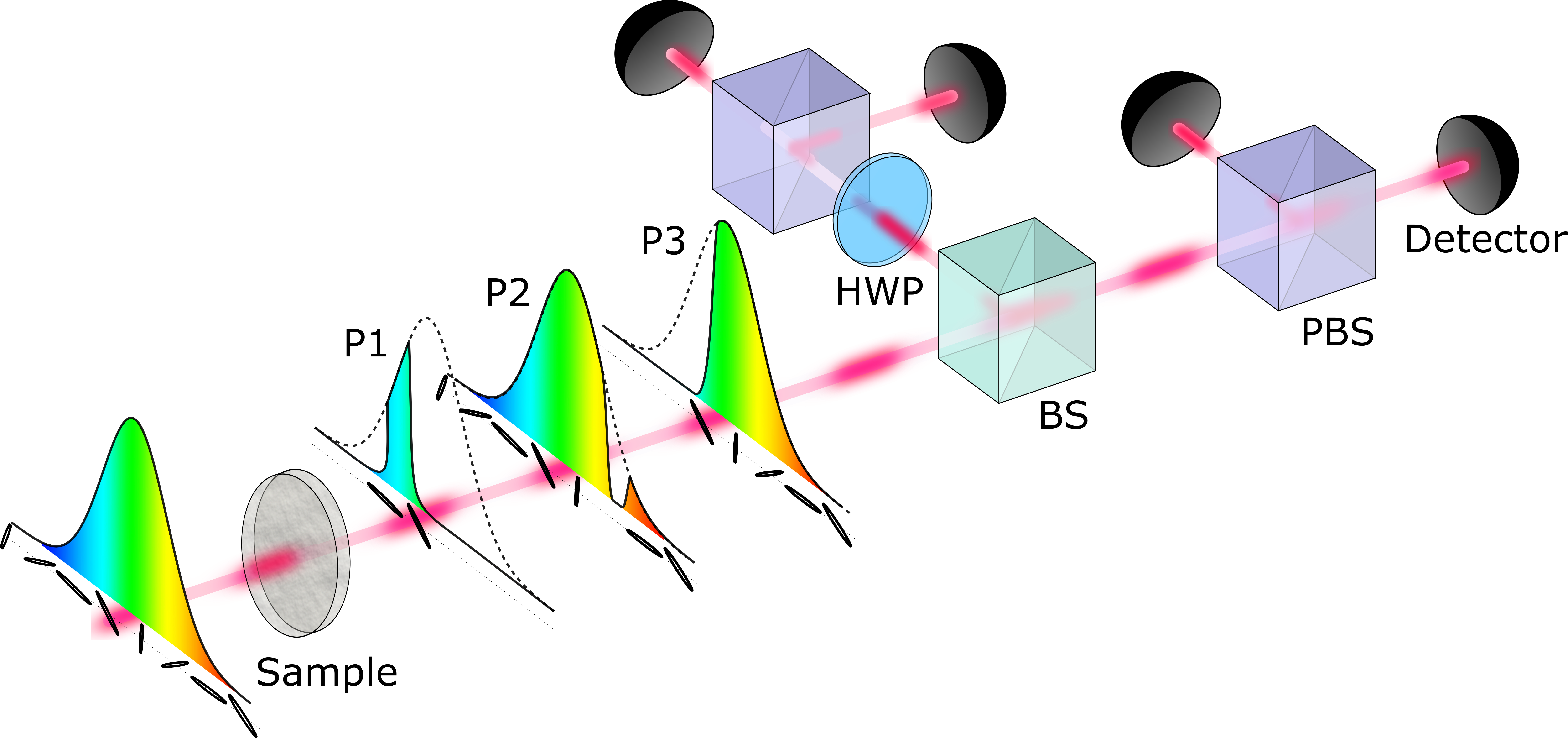}
\caption{Conceptual idea. 
A train of spectral vector pulses propagates through a sample, which could have different absorbing characteristics from pulse to pulse, such as a narrowband transmission (P1), a narrowband absorption (P2), or a long-pass filter (P3).
Subsequently, the linear polarization components of the pulses are analyzed using a beam splitter (BS), polarizing beam splitters (PBS), a half-wave plate (HWP), and photodetectors to reconstruct the perturbation of the spectrum.
The polarization pattern over wavelength is denoted beneath the spectrum.}
  \label{fig:plot1}
\end{figure}

Here, we extend these ideas to the spectral domain and generate states of light that possess a different polarization state for every wavelength. 
In analogy to the spatial domain, we term these states \textit{spectral vector beams} (SVBs).
We demonstrate that a simple modulation of a femtosecond laser pulse in the time domain using a birefringent crystal enables a controlled and flexible way of generating different polarization patterns across the frequency bandwidth of the light.
We further show that this correlation can be exploited in spectroscopic measurements using only polarization measurements.
Using fast photodiodes, we can track spectral modulations with read-out rates of up to 6\,MHz, which, in the current experiment, is mainly limited by the speed of our frequency modulation scheme. 
In general, the presented method is capable of tracking pulse-to-pulse variations in the frequency spectrum, such that high-speed spectroscopic measurements with GHz read-out rates are feasible with current technologies.
The general concept is sketched in Fig.~\ref{fig:plot1}.
Finally, we outline how the spectral measurement range can be extended using coherent supercontinuum light sources, which will allow spectroscopy over the whole NIR, or IR spectrum.

\section{Results}

\subsection{Spectral Vector Beams}

Beams having a varying polarization vector across their frequency spectrum, i.e.~light fields that have a strong correlation between frequency and polarization, we term spectral vector beam (SVB). 
In order to obtain a polarization state that continuously changes over the frequency spectrum, a light field is required to have a varying relative phase along the frequency spectrum between two orthogonal polarization components.
A simple way to realize a continuous phase shift in the frequency domain is to modulate the light in the temporal domain. 
Here, a time shift $\tau$ of the light field translates into a linear phase shift in the frequency domain.
Hence, coherently superimposing two orthogonally polarized pulses with a time delay $\tau$ gives rise to the wanted correlation between polarization and frequency, i.e.~a SVB.

In the following, we discuss, without loss of generality, SVBs for which a linear polarization changes its polarization angle across the frequency range.
To generate such beams, two electric fields with identical envelope $E(t)$ are coherently superimposed.
The two fields are circularly polarized, one being right-handed and the other left-handed, and delayed by $\tau$ with respect to each other.  
Using the Jones vector formalism, we can write
\begin{equation}
\vec{E}(t)=\begin{pmatrix} 1 \\ -i \end{pmatrix} E(t) +\begin{pmatrix} 1 \\ i \end{pmatrix} E(t+\tau).
\label{eq:Beam}
\end{equation}
By taking the Fourier transform and utilizing the shift theorem, the signal is transposed into the frequency domain.
\begin{equation}
\vec{E}(\omega)=\begin{pmatrix} 1 \\ -i \end{pmatrix} \mathcal{F}\left[E(t)\right] +\begin{pmatrix}1 \\ i \end{pmatrix} e^{-i\omega \tau} \mathcal{F}\left[ E(t)\right]
\label{eq:BeamF}
\end{equation}
Thus, the two orthogonal terms of Eq.~\ref{eq:Beam} have a linear phase shift of $\omega\tau$ leading to a rotating linear polarization over frequency, as illustrated in Fig.~\ref{fig:plot1}.
With an appropriate temporal delay, it is possible to obtain a SVB for which the polarization rotates precisely once over the spectrum.
In that case, each linear polarization angle corresponds to one frequency, such that a simple polarization measurement allows to draw conclusions about spectral changes. 
When averaging the signal over time and frequency, the light has a low degree of polarization.
For long time delays and nearly no temporal overlap between the two trailing components, the degree of polarization further decreases until it is seemingly unpolarized.
We note that the features of SVBs might thus also be interesting from a fundamental point of view. For example, they can be used to investigate analogies between classical and quantum optics \cite{spreeuw1998classical,aiello2015quantum,forbes2019classically} and to study geometric phases discussed in a similar context in other domains \cite{hannonen2019geometric,hannonen2020measurement}.

\subsection{Generating Spectral Vector Beams}

\begin{figure*}[htb] 
  \centering
  \includegraphics[width=0.95\textwidth]{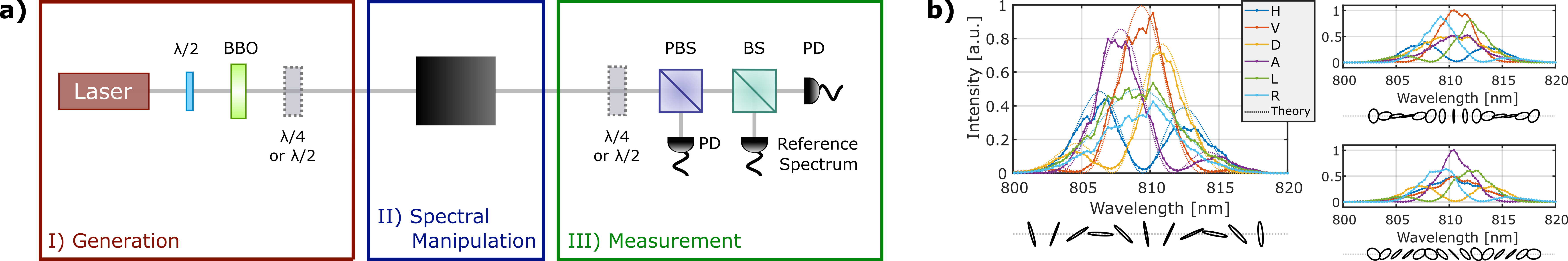}
  \caption{Experimental generation of SVBs. 
  a) Sketch of the experimental setup.
 I) SVBs are generated using a diagonally polarized pulse whose vertical polarized component is delayed by propagating through a birefringent crystal, such as a BaB$_2$O$_4$ (BBO) crystal.
 A quarter- or half-wave plate ($\lambda/4$ or $\lambda/2$) is used to realize the spectral polarization patterns displayed in b).
 II) The spectral components of the pulse are manipulated to simulate different absorption patterns of a sample to demonstrate spectroscopic measurements. 
 The experimental realisations are displayed in figures~\ref{fig:plot3} and \ref{fig:plot4}.
 III) A combination of wave plates and a polarizing beam splitter (PBS) together with photodiodes (PD) project the polarization state into different polarization bases.
 A reference measurement using another beam splitter (BS) and a spectrometer is used for reference and comparison.
 b) Measured spectra of three different SVBs with their corresponding polarization patterns, containing only linear polarisation (left) and different linear and circular polarization states (right).
 The legend denotes H, V, D, A, L, R for horizontal, vertical, diagonal, antidiagonal, and left- and right-hand circular polarizations, respectively.
 The spectra are generated using a quarter-wave plate at 45$^\circ$ between the fast and slow axis of the wave plate and the axis of polarization of the incident beam (left), half-wave plate at 22.5$^\circ$ (upper right), or no wave plate (lower right) after the BBO crystal. 
}
  \label{fig:plot2}
\end{figure*}

In the experiment, we use a pulsed Ti:Sapphire laser with a pulse duration of around 220\,fs  and an 80\,MHz repetition rate centered around a wavelength of 808\,nm.
The pulse is diagonally polarized with respect to the optical axis of a birefringent BaB$_2$O$_4$ (BBO) crystal.
The crystal has different group indices for the horizontally and vertically polarized components $\Delta n_g=-0.12$. 
Therefore, propagation through the BBO crystal delays the vertically polarized part. 
A combination of wave plates is then used to convert the wavelength-dependent polarization change to other polarization bases.
A sketch of the experimental setup is shown in Fig.~\ref{fig:plot2}\,a. 

In the first set of experiments, we demonstrate the generation of different SVBs by performing a spectrally-resolved polarization tomography using polarization optics and a spectrometer. 
Hereby, we utilize a 2\,mm thick crystal in order to achieve a time delay of roughly the pulse length ($\tau\approx 220$\,fs), which results in a $2\pi$ phase ramp over the spectrum of the pulse.
The different polarization components of the spectrum are measured with our spectrometer and are used to deduce the polarization states for each wavelength.
When superposing two circularly-polarized trailing pulses (as described in Eq. \eqref{eq:BeamF}), a linear polarization that rotates over the spectrum is generated, as shown in Fig.~\ref{fig:plot2}\,b on the left.
SVBs with different polarization patterns can be generated by using different waveplate orientations, as shown in Fig.~\ref{fig:plot2}\,b on the right.
As described earlier, if the temporal and frequency structure is ignored, these beams have a low overall degree of polarization ($p$). 
Experimentally, we obtain values $p\approx$\,0.25 for the beams presented in Fig.~\ref{fig:plot2}\,b. 
The obtained pattern can be further tuned, e.g.~by using crystals of different thicknesses, thereby changing the delay between the two parts of the pulse and the slope of the rotation angle (see supplementary). 
When using SVBs for sensing, this tuning can be utilized to increase the spectral resolution at the cost of possible ambiguities on the wavelength range.
Additionally, the obtained polarization patterns can be spectrally shifted by adding a relative phase between the two superposed parts of the pulse, e.g.~through a slight tilt of the birefringent crystal (see supplementary).

\subsection{Spectroscopic measurements}

To perform spectroscopic measurements, we use SVBs as described by Eq.~\eqref{eq:BeamF} and shown in Fig.~\ref{fig:plot2}\,b on the left side.
We track different spectral manipulations by measuring the change of polarization with only the photodetectors, from which the Stokes parameters are readily obtained (see supplementary).
At first, the system is calibrated by measuring the Stokes parameters $S_1$ and $S_2$ across the whole wavelength range for the unperturbed spectrum with a spectrometer. 
Then, the crystal width, pulse width, and the phase used in the theoretical simulations is adjusted in order to match the theoretically predicted spectrum with the experimental one.
From the optimized simulations, we can deduce the polarization state over wavelength.
Hereby, the rotation angle $\theta$ of the polarization ellipse is pivotal.
It relates $S_1$ and $S_2$ and is given by
\begin{equation}
\theta=\frac{1}{2}\arctan\left(\frac{S_2}{S_1}\right).
\end{equation}
Knowing the exact relation between $\theta$ and the wavelength allows to deduce the wavelength by comparing the measured rotation angle with the simulated predictions. 
We note that for an ideal preparation of this state, the circular polarisation components summed up in $S_3$ always add up to zero. 
In experiments, imperfect optical elements might lead to a nonzero $S_3$ value (see supplementary). 
However, they are not considered in the analysis since they do not contribute to the determination of the rotation angle.
The polarization measurements are done using silicon photodetectors with a bandwidth of 350\,MHz and 50\,ps rise times.
In general, the simultaneous projections into two linear polarization bases are necessary to obtain the required Stokes parameters in a single shot measurement (see the sketch in Fig.~\ref{fig:plot1}).
In our experiment, we perform the projections into the two bases consecutively by adjusting the half-wave plate in the polarization analysis setup accordingly (see the setup in Fig.~\ref{fig:plot2}).

\begin{figure*}[ht] 
  \centering
  \includegraphics[width=0.9\textwidth]{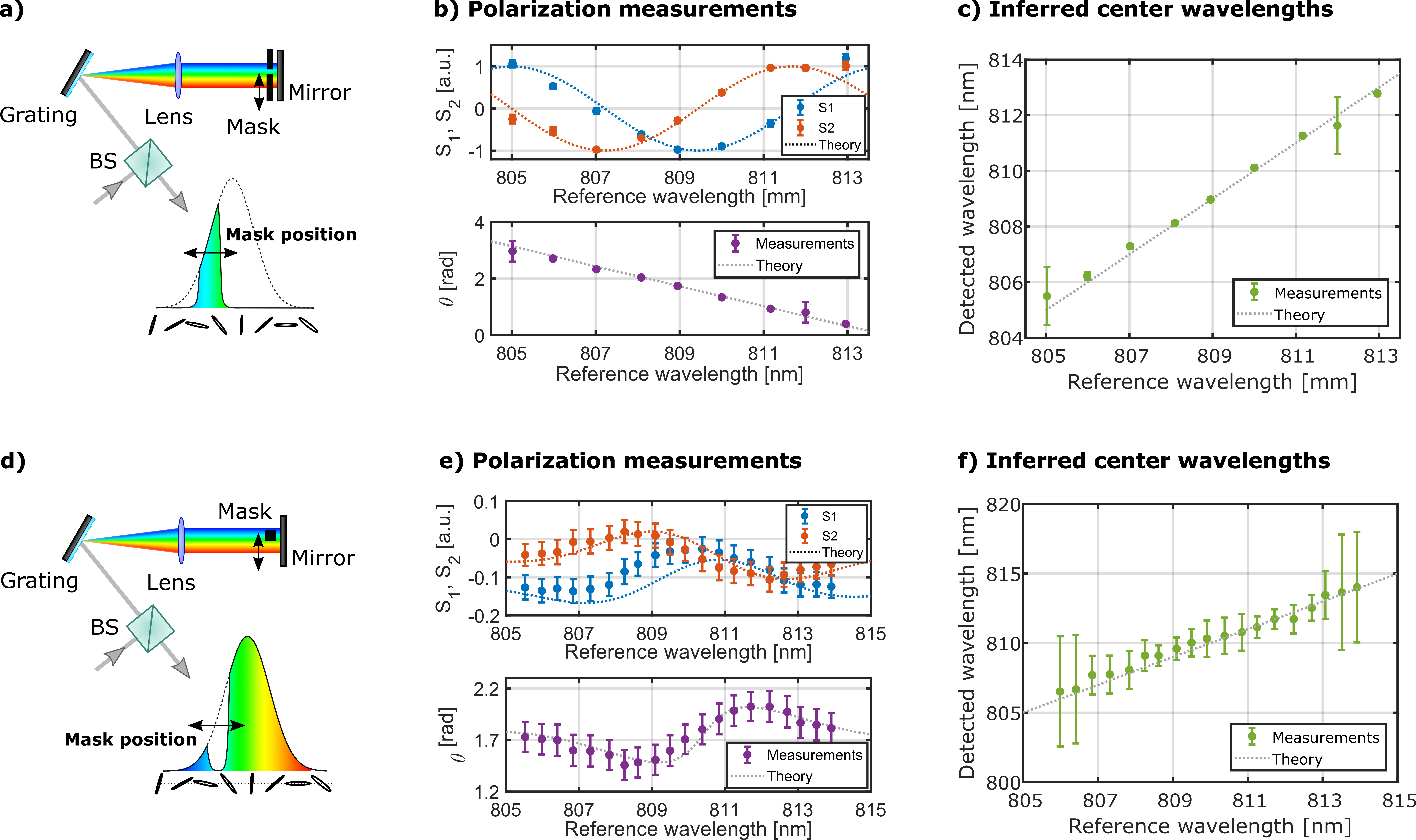}
  \caption{Spectroscopic measurements using SVBs.
  a) Setup to realise the transmission of a narrow wavelength band, in which a slit is moved through the spectral Fourier plane of the light field.
  b) The measured Stokes parameters and the corresponding rotation angle $\theta$ are derived from the photodetector measurements are plotted over the reference wavelength measured with the spectrometer.
  c) The center wavelength of the band-pass filter is inferred from the characterization of the polarization and compared to the reference wavelength.
  d) Setup to realise a narrowband wavelength absorption of the spectrum, implemented by placing a wire in the spectral Fourier plane.
  e) The measured Stokes parameters and $\theta$ as a function of the reference wavelength.
  f) The inferred center wavelengths of the narrowband absorption corresponding to the expected values.
  All errors are obtained by taking the standard deviation of multiple consecutive measurements with the same settings.
}
  \label{fig:plot3}
\end{figure*}

\begin{figure}[ht] 
  \centering
  \includegraphics[width=\linewidth]{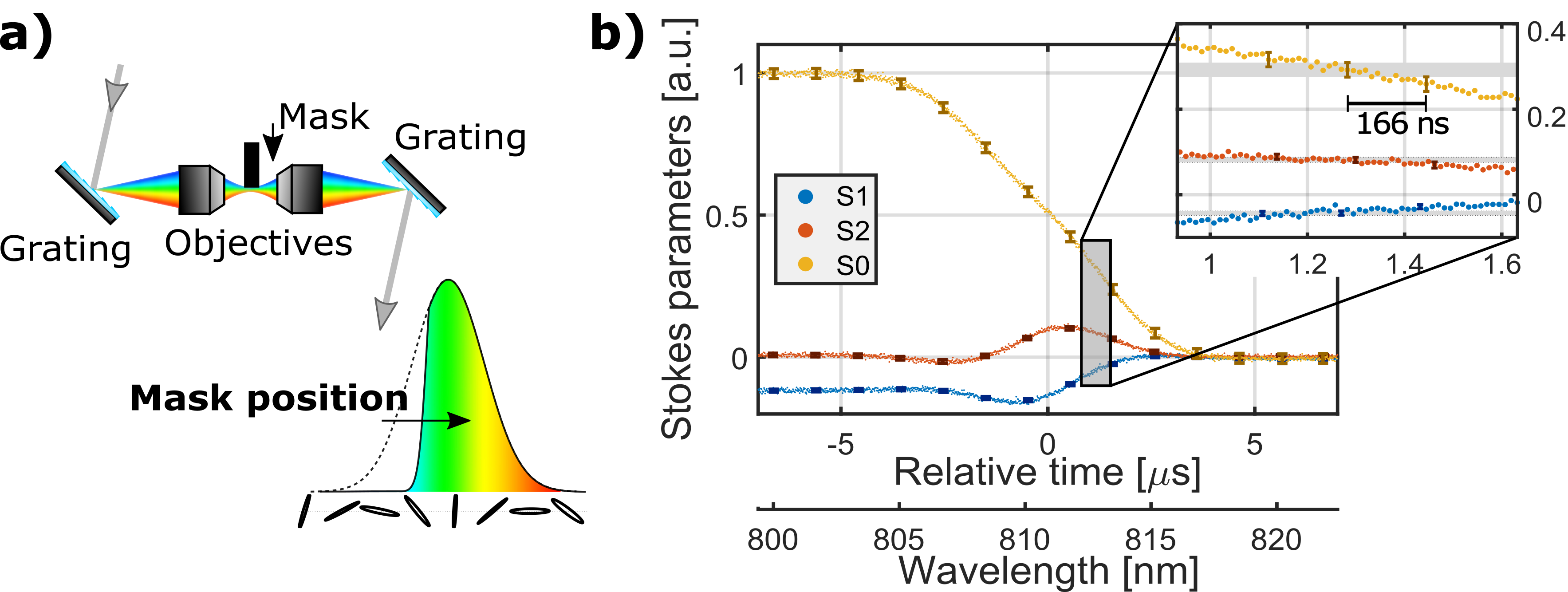}
  \caption{High-speed tracking of frequency modulations. 
  a) The setup for high-speed measurements. 
  A chopper wheel blade is placed in the Fourier plane of a $4f$-system composed of diffraction gratings and two 20$\times$ microscope objectives. 
  b) The Stokes parameter measurement tracks the change in polarization, which can be correlated to the absorbed wavelength components. 
  Modulations down to 166.5\,ns are resolved within the bounds of the errors.}
  \label{fig:plot4}
\end{figure}

To demonstrate the broad range of applicabilities of this method, three different spectral manipulations are simulated: the transmission of a narrow band (P1), the absorption of a narrow band (P2), and a fast varying long-pass filter (P3). 
These three different measurement scenarios are sketched in Fig.~\ref{fig:plot1}.
In every approach, the frequency spectrum of the SVB is manipulated by placing the corresponding absorbing masks into the spectral Fourier plane of the beam obtained using the experimental settings depicted on the left in Figs.~\ref{fig:plot3} and \ref{fig:plot4}.

First, we show a polarization measurement tracking the change of the center wavelength of a narrowband transmission.
A 210\,$\mu$m broad slit is placed into the spectral Fourier plane of the pulse as shown in Fig.~\ref{fig:plot3}\,a. 
The movable slit generates a transmission bandwidth of 0.76\,nm (FWHM) 
in the frequency domain, which is on the verge of the resolution of the reference spectrometer specified as 0.7\,nm.
Different positions of the slit correspond to different center wavelengths of the bandpass filter with invariable bandwidth.
Only the polarization components within the transmission bandwidth appear in the subsequent polarization measurements shown in Fig.~\ref{fig:plot3}\,b.
By comparing the measured rotation angle with the simulated predictions based on the calibration spectrum, the center wavelength of the filter is inferred (see Fig.~\ref{fig:plot3}\,c).
The transmission measurements were obtained by averaging up to 80 pulses and have a standard deviation of 0.30\,nm averaged over all reconstructed wavelengths. 
The mean deviation between the reference wavelengths and the inferred center wavelengths is 0.20\,nm.  
The errors are mainly caused by pulse to pulse intensity fluctuations of our fs-laser.
Averaging over several pulses can thus greatly increase the precision at the cost of speed.
The narrowband filtering causes a significant increase of $\Delta p=0.67$ to $p=0.94\pm$0.15 averaged over all measurements. 
We note, that the knowledge of $p$ could also be used to deduce the bandwidth of the filter (see supplementary).

In the second set of measurements, a movable wire is placed in the spectral Fourier plane replicating an absorption line shifting across a wavelength band (see Fig.~\ref{fig:plot3}\,d).
Again, this modulation, i.e. its center wavelength, is tracked through measuring the polarization of the remaining light using Stokes parameters.
From these values, we deduce the center wavelength of the absorption line.
The wire has a diameter of 0.26\,mm, mimicking a tunable absorption band with a bandwidth of 1.17\,nm in the frequency domain. 
The absorption is tracked with an averaged standard deviation of 1.84\,nm (max. 800 pulses per wavelength), while the mean deviation between the expected wavelength and inferred center wavelengths is 0.34\,nm. 
The larger errors compared to the transmission measurements are caused by the more complex behavior of $\theta$.
An additional ambiguity and errors at the extrema of the function (see Fig.~\ref{fig:plot3}\,e) reduce the precision of the measurement.
Finally, the narrowband absorption changes the overall $p$ only slightly by maximally $\Delta p=0.07$, which shows another challenge in tracking these small changes in the polarization state.

In the final set of experiments, we show the high-speed capabilities of this technique using the setup shown in Fig.~\ref{fig:plot4}\,a. 
Here, a chopper wheel blade is placed in the spectral Fourier plane of a $4f$-system composed of two 20$\times$ microscope objectives.
The mask resembles a time-varying longpass filter resulting in a constant increase of $p$ and a constant decrease of the overall intensity. 
Changes in the spectrum down to 166.5\,ns (6\,MHz) are resolved within the bounds of the errors. 
The errors could be significantly reduced with a more stable movable block and a more stable light source. 
In our implementation, the measuring speed is mainly limited by the speed of our frequency modulation scheme, i.e. a speed of 25.6\,m/s with which the chopper wheel passes through the spectrum.
Nevertheless, this demonstration shows the potential to measure changes in wavelength with high-speed read-out rates, which in principle is only limited by the repetition rate of the laser and the rise time of the electronics.
Thus, spectroscopic measurements with GHz read-out rates seem accessible when using SVBs.

\subsection{Supercontinuum Spectral Vector Beams}

One constraint of the current experimental demonstration is the limited spectroscopic range when using ultra-short laser pulses.
However, since the presented method in principle works with any pulse shape, broadband applications could utilize coherent supercontinuum light sources \cite{Dudley2006Supercontinuum}.
Simulations using a standard supercontinuum light field suggest the possibility to cover wavelengths from 1000 to 2300\,nm ($-40$\,dB bandwidth, see the methods section). 
Hereby, the spectrum is polarized and superimposed with its orthogonally polarized and delayed counterpart.
A delay of 5.5\,fs hereby generates one full $\pi$ rotation of the polarization across the frequency spectrum (see the supplementary material).
With a similar performance as the presented measurements, a resolution of 43\,nm seems feasible.
The simulated results are shown and discussed in more detail in the supplementary material.
Since most materials have a significant birefringence dispersion over such broad wavelength ranges, simply using a birefringent crystal to generate an SVB will result in a complex spectral polarization pattern.
Alternatively, using an unbalanced interferometer to generate two orthogonally polarized supercontinuum fields with a slight delay would lead to similarly simple polarization patterns.\\

\section{Discussion}

In summary, we demonstrate a simple method to generate spectral vector beams, i.e.~light that has a strong correlation between its polarization and wavelength components.
Benefiting from this direct correlation, we show that high-speed spectroscopy, e.g. tracking a narrow line absorption or transmission, can be done using only polarization measurements.
Because such measurements can be performed with fast photodetectors, the method is capable of tracking changes of the frequency spectrum with read-out rates that are only limited by the repetition rate of the laser and the response time of the detectors, i.e.~in the GHz regime.
We further outline a way to overcome the limited wavelength range of our light source and describe that a similar method could be applied for broadband spectroscopic sensing using a supercontinuum source.
Besides the experimental implementation of a supercontinuum SVB, it will be interesting to also study broadband incoherent light sources such as LEDs in a similar setting in the future. Thereby, the effect of spectral coherence and low energy densities on SVBs and the presented high-speed spectroscopic method can be investigated.
Additionally, it will be important to study extensions of the presented approach to enable the tracking of more complex modulations of the spectrum, e.g.~through generating more complex polarization patterns and implementing more advanced signal processing schemes.
Finally, we anticipate that the benefits of the simple generation of SVBs, as well as the spectroscopic method demonstrated here, will stimulate further research into the complex structuring of light harnessing the spectral and polarization degree of freedom of light.


\begin{thebibliography}{24}%
\makeatletter
\providecommand \@ifxundefined [1]{%
 \@ifx{#1\undefined}
}%
\providecommand \@ifnum [1]{%
 \ifnum #1\expandafter \@firstoftwo
 \else \expandafter \@secondoftwo
 \fi
}%
\providecommand \@ifx [1]{%
 \ifx #1\expandafter \@firstoftwo
 \else \expandafter \@secondoftwo
 \fi
}%
\providecommand \natexlab [1]{#1}%
\providecommand \enquote  [1]{``#1''}%
\providecommand \bibnamefont  [1]{#1}%
\providecommand \bibfnamefont [1]{#1}%
\providecommand \citenamefont [1]{#1}%
\providecommand \href@noop [0]{\@secondoftwo}%
\providecommand \href [0]{\begingroup \@sanitize@url \@href}%
\providecommand \@href[1]{\@@startlink{#1}\@@href}%
\providecommand \@@href[1]{\endgroup#1\@@endlink}%
\providecommand \@sanitize@url [0]{\catcode `\\12\catcode `\$12\catcode
  `\&12\catcode `\#12\catcode `\^12\catcode `\_12\catcode `\%12\relax}%
\providecommand \@@startlink[1]{}%
\providecommand \@@endlink[0]{}%
\providecommand \url  [0]{\begingroup\@sanitize@url \@url }%
\providecommand \@url [1]{\endgroup\@href {#1}{\urlprefix }}%
\providecommand \urlprefix  [0]{URL }%
\providecommand \Eprint [0]{\href }%
\providecommand \doibase [0]{http://dx.doi.org/}%
\providecommand \selectlanguage [0]{\@gobble}%
\providecommand \bibinfo  [0]{\@secondoftwo}%
\providecommand \bibfield  [0]{\@secondoftwo}%
\providecommand \translation [1]{[#1]}%
\providecommand \BibitemOpen [0]{}%
\providecommand \bibitemStop [0]{}%
\providecommand \bibitemNoStop [0]{.\EOS\space}%
\providecommand \EOS [0]{\spacefactor3000\relax}%
\providecommand \BibitemShut  [1]{\csname bibitem#1\endcsname}%
\let\auto@bib@innerbib\@empty
\bibitem [{\citenamefont {Gauglitz}\ and\ \citenamefont
  {Moore}(2014)}]{gauglitz2014handbook}%
  \BibitemOpen
  \bibfield  {author} {\bibinfo {author} {\bibfnamefont {G{\"u}nter}\
  \bibnamefont {Gauglitz}}\ and\ \bibinfo {author} {\bibfnamefont
  {David~Steven}\ \bibnamefont {Moore}},\ }\href@noop {} {\emph {\bibinfo
  {title} {Handbook of spectroscopy}}},\ Vol.~\bibinfo {volume} {1}\ (\bibinfo
  {publisher} {Wiley Online Library},\ \bibinfo {year} {2014})\BibitemShut
  {NoStop}%
\bibitem [{\citenamefont {Tkachenko}(2006)}]{tkachenko2006optical}%
  \BibitemOpen
  \bibfield  {author} {\bibinfo {author} {\bibfnamefont {Nikolai~V}\
  \bibnamefont {Tkachenko}},\ }\href@noop {} {\emph {\bibinfo {title} {Optical
  spectroscopy: methods and instrumentations}}}\ (\bibinfo  {publisher}
  {Elsevier},\ \bibinfo {year} {2006})\BibitemShut {NoStop}%
\bibitem [{\citenamefont {Weiner}(2011)}]{weiner2011ultrafast}%
  \BibitemOpen
  \bibfield  {author} {\bibinfo {author} {\bibfnamefont {Andrew}\ \bibnamefont
  {Weiner}},\ }\bibfield  {title} {\enquote {\bibinfo {title} {Ultrafast
  optical pulse shaping: A tutorial review},}\ }\href {\doibase
  10.1016/j.optcom.2011.03.084} {\bibfield  {journal} {\bibinfo  {journal}
  {Optics Communications - OPT COMMUN}\ }\textbf {\bibinfo {volume} {284}},\
  \bibinfo {pages} {3669--3692} (\bibinfo {year} {2011})}\BibitemShut {NoStop}%
\bibitem [{\citenamefont {Rubinsztein-Dunlop}\ \emph
  {et~al.}(2016)\citenamefont {Rubinsztein-Dunlop}, \citenamefont {Forbes},
  \citenamefont {Berry}, \citenamefont {Dennis}, \citenamefont {Andrews},
  \citenamefont {Mansuripur}, \citenamefont {Denz}, \citenamefont {Alpmann},
  \citenamefont {Banzer}, \citenamefont {Bauer} \emph
  {et~al.}}]{rubinsztein2016roadmap}%
  \BibitemOpen
  \bibfield  {author} {\bibinfo {author} {\bibfnamefont {Halina}\ \bibnamefont
  {Rubinsztein-Dunlop}}, \bibinfo {author} {\bibfnamefont {Andrew}\
  \bibnamefont {Forbes}}, \bibinfo {author} {\bibfnamefont {Michael~V}\
  \bibnamefont {Berry}}, \bibinfo {author} {\bibfnamefont {Mark~R}\
  \bibnamefont {Dennis}}, \bibinfo {author} {\bibfnamefont {David~L}\
  \bibnamefont {Andrews}}, \bibinfo {author} {\bibfnamefont {Masud}\
  \bibnamefont {Mansuripur}}, \bibinfo {author} {\bibfnamefont {Cornelia}\
  \bibnamefont {Denz}}, \bibinfo {author} {\bibfnamefont {Christina}\
  \bibnamefont {Alpmann}}, \bibinfo {author} {\bibfnamefont {Peter}\
  \bibnamefont {Banzer}}, \bibinfo {author} {\bibfnamefont {Thomas}\
  \bibnamefont {Bauer}},  \emph {et~al.},\ }\bibfield  {title} {\enquote
  {\bibinfo {title} {Roadmap on structured light},}\ }\href {\doibase
  https://doi.org/10.1088/2040-8978/19/1/013001} {\bibfield  {journal}
  {\bibinfo  {journal} {Journal of Optics}\ }\textbf {\bibinfo {volume} {19}},\
  \bibinfo {pages} {013001} (\bibinfo {year} {2016})}\BibitemShut {NoStop}%
\bibitem [{\citenamefont {Walecki}\ \emph {et~al.}(1997)\citenamefont
  {Walecki}, \citenamefont {Fittinghoff}, \citenamefont {Smirl},\ and\
  \citenamefont {Trebino}}]{Walecki1997Characterization}%
  \BibitemOpen
  \bibfield  {author} {\bibinfo {author} {\bibfnamefont {W.~J.}\ \bibnamefont
  {Walecki}}, \bibinfo {author} {\bibfnamefont {David~N.}\ \bibnamefont
  {Fittinghoff}}, \bibinfo {author} {\bibfnamefont {Arthur~L.}\ \bibnamefont
  {Smirl}}, \ and\ \bibinfo {author} {\bibfnamefont {Rick}\ \bibnamefont
  {Trebino}},\ }\bibfield  {title} {\enquote {\bibinfo {title}
  {Characterization of the polarization state of weak ultrashort coherent
  signals by dual-channel spectral interferometry},}\ }\href {\doibase
  10.1364/OL.22.000081} {\bibfield  {journal} {\bibinfo  {journal} {Opt.
  Lett.}\ }\textbf {\bibinfo {volume} {22}},\ \bibinfo {pages} {81--83}
  (\bibinfo {year} {1997})}\BibitemShut {NoStop}%
\bibitem [{\citenamefont {Villeneuve}\ \emph {et~al.}(2000)\citenamefont
  {Villeneuve}, \citenamefont {Aseyev}, \citenamefont {Dietrich}, \citenamefont
  {Spanner}, \citenamefont {Ivanov},\ and\ \citenamefont
  {Corkum}}]{Villeneuve2000Forced}%
  \BibitemOpen
  \bibfield  {author} {\bibinfo {author} {\bibfnamefont {D.~M.}\ \bibnamefont
  {Villeneuve}}, \bibinfo {author} {\bibfnamefont {S.~A.}\ \bibnamefont
  {Aseyev}}, \bibinfo {author} {\bibfnamefont {P.}~\bibnamefont {Dietrich}},
  \bibinfo {author} {\bibfnamefont {M.}~\bibnamefont {Spanner}}, \bibinfo
  {author} {\bibfnamefont {M.~Yu.}\ \bibnamefont {Ivanov}}, \ and\ \bibinfo
  {author} {\bibfnamefont {P.~B.}\ \bibnamefont {Corkum}},\ }\bibfield  {title}
  {\enquote {\bibinfo {title} {Forced molecular rotation in an optical
  centrifuge},}\ }\href {\doibase 10.1103/PhysRevLett.85.542} {\bibfield
  {journal} {\bibinfo  {journal} {Phys. Rev. Lett.}\ }\textbf {\bibinfo
  {volume} {85}},\ \bibinfo {pages} {542--545} (\bibinfo {year}
  {2000})}\BibitemShut {NoStop}%
\bibitem [{\citenamefont {Koehl}\ \emph {et~al.}(1998)\citenamefont {Koehl},
  \citenamefont {Hattori},\ and\ \citenamefont {Nelson}}]{koehl1998automated}%
  \BibitemOpen
  \bibfield  {author} {\bibinfo {author} {\bibfnamefont {Richard~M.}\
  \bibnamefont {Koehl}}, \bibinfo {author} {\bibfnamefont {Toshiaki}\
  \bibnamefont {Hattori}}, \ and\ \bibinfo {author} {\bibfnamefont {Keith~A.}\
  \bibnamefont {Nelson}},\ }\bibfield  {title} {\enquote {\bibinfo {title}
  {Automated spatial and temporal shaping of femtosecond pulses},}\ }\href
  {\doibase https://doi.org/10.1016/S0030-4018(98)00486-6} {\bibfield
  {journal} {\bibinfo  {journal} {Optics Communications}\ }\textbf {\bibinfo
  {volume} {157}},\ \bibinfo {pages} {57 -- 61} (\bibinfo {year}
  {1998})}\BibitemShut {NoStop}%
\bibitem [{\citenamefont {Shaltout}\ \emph {et~al.}(2019)\citenamefont
  {Shaltout}, \citenamefont {Lagoudakis}, \citenamefont {van~de Groep},
  \citenamefont {Kim}, \citenamefont {Vu{\v{c}}kovi{\'c}}, \citenamefont
  {Shalaev},\ and\ \citenamefont {Brongersma}}]{shaltout2019spatiotemporal}%
  \BibitemOpen
  \bibfield  {author} {\bibinfo {author} {\bibfnamefont {Amr~M}\ \bibnamefont
  {Shaltout}}, \bibinfo {author} {\bibfnamefont {Konstantinos~G}\ \bibnamefont
  {Lagoudakis}}, \bibinfo {author} {\bibfnamefont {Jorik}\ \bibnamefont {van~de
  Groep}}, \bibinfo {author} {\bibfnamefont {Soo~Jin}\ \bibnamefont {Kim}},
  \bibinfo {author} {\bibfnamefont {Jelena}\ \bibnamefont
  {Vu{\v{c}}kovi{\'c}}}, \bibinfo {author} {\bibfnamefont {Vladimir~M}\
  \bibnamefont {Shalaev}}, \ and\ \bibinfo {author} {\bibfnamefont {Mark~L}\
  \bibnamefont {Brongersma}},\ }\bibfield  {title} {\enquote {\bibinfo {title}
  {Spatiotemporal light control with frequency-gradient metasurfaces},}\ }\href
  {\doibase 10.1126/science.aax2357} {\bibfield  {journal} {\bibinfo  {journal}
  {Science}\ }\textbf {\bibinfo {volume} {365}},\ \bibinfo {pages} {374--377}
  (\bibinfo {year} {2019})}\BibitemShut {NoStop}%
\bibitem [{\citenamefont {Forbes}\ \emph {et~al.}(2016)\citenamefont {Forbes},
  \citenamefont {Dudley},\ and\ \citenamefont {McLaren}}]{forbes2016creation}%
  \BibitemOpen
  \bibfield  {author} {\bibinfo {author} {\bibfnamefont {Andrew}\ \bibnamefont
  {Forbes}}, \bibinfo {author} {\bibfnamefont {Angela}\ \bibnamefont {Dudley}},
  \ and\ \bibinfo {author} {\bibfnamefont {Melanie}\ \bibnamefont {McLaren}},\
  }\bibfield  {title} {\enquote {\bibinfo {title} {Creation and detection of
  optical modes with spatial light modulators},}\ }\href {\doibase
  10.1364/AOP.8.000200} {\bibfield  {journal} {\bibinfo  {journal} {Adv. Opt.
  Photon.}\ }\textbf {\bibinfo {volume} {8}},\ \bibinfo {pages} {200--227}
  (\bibinfo {year} {2016})}\BibitemShut {NoStop}%
\bibitem [{\citenamefont {Pinnegar}(2006)}]{Pinnegar2006Polarization}%
  \BibitemOpen
  \bibfield  {author} {\bibinfo {author} {\bibfnamefont {Rob}\ \bibnamefont
  {Pinnegar}},\ }\bibfield  {title} {\enquote {\bibinfo {title} {Polarization
  analysis and polarization filtering of three-component signals with the
  time-frequency s transform},}\ }\href {\doibase
  10.1111/j.1365-246X.2006.02937.x} {\bibfield  {journal} {\bibinfo  {journal}
  {Geophysical Journal International}\ }\textbf {\bibinfo {volume} {165}},\
  \bibinfo {pages} {596--606} (\bibinfo {year} {2006})}\BibitemShut {NoStop}%
\bibitem [{\citenamefont {Fade}\ and\ \citenamefont
  {Alouini}(2012)}]{Fade2012Depolarization}%
  \BibitemOpen
  \bibfield  {author} {\bibinfo {author} {\bibfnamefont {Julien}\ \bibnamefont
  {Fade}}\ and\ \bibinfo {author} {\bibfnamefont {Mehdi}\ \bibnamefont
  {Alouini}},\ }\bibfield  {title} {\enquote {\bibinfo {title} {Depolarization
  remote sensing by orthogonality breaking},}\ }\href {\doibase
  10.1103/PhysRevLett.109.043901} {\bibfield  {journal} {\bibinfo  {journal}
  {Phys. Rev. Lett.}\ }\textbf {\bibinfo {volume} {109}},\ \bibinfo {pages}
  {043901} (\bibinfo {year} {2012})}\BibitemShut {NoStop}%
\bibitem [{\citenamefont {Sano}\ \emph {et~al.}(1980)\citenamefont {Sano},
  \citenamefont {Okada},\ and\ \citenamefont {Hashimoto}}]{Sano1980Simple}%
  \BibitemOpen
  \bibfield  {author} {\bibinfo {author} {\bibfnamefont {K.}~\bibnamefont
  {Sano}}, \bibinfo {author} {\bibfnamefont {K.}~\bibnamefont {Okada}}, \ and\
  \bibinfo {author} {\bibfnamefont {K.}~\bibnamefont {Hashimoto}},\ }\bibfield
  {title} {\enquote {\bibinfo {title} {Simple optical wavelength meter in
  700–1200 nm wavelength region},}\ }\href
  {https://digital-library.theiet.org/content/journals/10.1049/el_19800650}
  {\bibfield  {journal} {\bibinfo  {journal} {Electronics Letters}\ }\textbf
  {\bibinfo {volume} {16}},\ \bibinfo {pages} {912--913(1)} (\bibinfo {year}
  {1980})}\BibitemShut {NoStop}%
\bibitem [{\citenamefont {Dimmick}(1997)}]{Dimmick1997Simple}%
  \BibitemOpen
  \bibfield  {author} {\bibinfo {author} {\bibfnamefont {Timothy~E.}\
  \bibnamefont {Dimmick}},\ }\bibfield  {title} {\enquote {\bibinfo {title}
  {Simple and accurate wavemeter implemented with a polarization
  interferometer},}\ }\href {\doibase 10.1364/AO.36.009396} {\bibfield
  {journal} {\bibinfo  {journal} {Appl. Opt.}\ }\textbf {\bibinfo {volume}
  {36}},\ \bibinfo {pages} {9396--9401} (\bibinfo {year} {1997})}\BibitemShut
  {NoStop}%
\bibitem [{\citenamefont {Dennis}\ \emph {et~al.}(2009)\citenamefont {Dennis},
  \citenamefont {O'Holleran},\ and\ \citenamefont
  {Padgett}}]{dennis2009singular}%
  \BibitemOpen
  \bibfield  {author} {\bibinfo {author} {\bibfnamefont {Mark~R.}\ \bibnamefont
  {Dennis}}, \bibinfo {author} {\bibfnamefont {Kevin}\ \bibnamefont
  {O'Holleran}}, \ and\ \bibinfo {author} {\bibfnamefont {Miles~J.}\
  \bibnamefont {Padgett}},\ }\bibfield  {title} {\enquote {\bibinfo {title}
  {Chapter 5 singular optics: Optical vortices and polarization
  singularities},}\ }\href {\doibase
  https://doi.org/10.1016/S0079-6638(08)00205-9} {\ \bibinfo {series} {Progress
  in Optics},\ \textbf {\bibinfo {volume} {53}},\ \bibinfo {pages} {293 -- 363}
  (\bibinfo {year} {2009})}\BibitemShut {NoStop}%
\bibitem [{\citenamefont {Rosales-Guzm{\'a}n}\ \emph
  {et~al.}(2018)\citenamefont {Rosales-Guzm{\'a}n}, \citenamefont {Ndagano},\
  and\ \citenamefont {Forbes}}]{rosales2018review}%
  \BibitemOpen
  \bibfield  {author} {\bibinfo {author} {\bibfnamefont {Carmelo}\ \bibnamefont
  {Rosales-Guzm{\'a}n}}, \bibinfo {author} {\bibfnamefont {Bienvenu}\
  \bibnamefont {Ndagano}}, \ and\ \bibinfo {author} {\bibfnamefont {Andrew}\
  \bibnamefont {Forbes}},\ }\bibfield  {title} {\enquote {\bibinfo {title} {A
  review of complex vector light fields and their applications},}\ }\href
  {\doibase https://doi.org/10.1088/2040-8986/aaeb7d} {\bibfield  {journal}
  {\bibinfo  {journal} {Journal of Optics}\ }\textbf {\bibinfo {volume} {20}},\
  \bibinfo {pages} {123001} (\bibinfo {year} {2018})}\BibitemShut {NoStop}%
\bibitem [{\citenamefont {Bauer}\ \emph {et~al.}(2015)\citenamefont {Bauer},
  \citenamefont {Banzer}, \citenamefont {Karimi}, \citenamefont {Orlov},
  \citenamefont {Rubano}, \citenamefont {Marrucci}, \citenamefont {Santamato},
  \citenamefont {Boyd},\ and\ \citenamefont {Leuchs}}]{bauer2015observation}%
  \BibitemOpen
  \bibfield  {author} {\bibinfo {author} {\bibfnamefont {Thomas}\ \bibnamefont
  {Bauer}}, \bibinfo {author} {\bibfnamefont {Peter}\ \bibnamefont {Banzer}},
  \bibinfo {author} {\bibfnamefont {Ebrahim}\ \bibnamefont {Karimi}}, \bibinfo
  {author} {\bibfnamefont {Sergej}\ \bibnamefont {Orlov}}, \bibinfo {author}
  {\bibfnamefont {Andrea}\ \bibnamefont {Rubano}}, \bibinfo {author}
  {\bibfnamefont {Lorenzo}\ \bibnamefont {Marrucci}}, \bibinfo {author}
  {\bibfnamefont {Enrico}\ \bibnamefont {Santamato}}, \bibinfo {author}
  {\bibfnamefont {Robert~W}\ \bibnamefont {Boyd}}, \ and\ \bibinfo {author}
  {\bibfnamefont {Gerd}\ \bibnamefont {Leuchs}},\ }\bibfield  {title} {\enquote
  {\bibinfo {title} {Observation of optical polarization m{\"o}bius strips},}\
  }\href {\doibase https://doi.org/10.1126/science.1260635} {\bibfield
  {journal} {\bibinfo  {journal} {Science}\ }\textbf {\bibinfo {volume}
  {347}},\ \bibinfo {pages} {964--966} (\bibinfo {year} {2015})}\BibitemShut
  {NoStop}%
\bibitem [{\citenamefont {Larocque}\ \emph {et~al.}(2018)\citenamefont
  {Larocque}, \citenamefont {Sugic}, \citenamefont {Mortimer}, \citenamefont
  {Taylor}, \citenamefont {Fickler}, \citenamefont {Boyd}, \citenamefont
  {Dennis},\ and\ \citenamefont {Karimi}}]{larocque2018reconstructing}%
  \BibitemOpen
  \bibfield  {author} {\bibinfo {author} {\bibfnamefont {Hugo}\ \bibnamefont
  {Larocque}}, \bibinfo {author} {\bibfnamefont {Danica}\ \bibnamefont
  {Sugic}}, \bibinfo {author} {\bibfnamefont {Dominic}\ \bibnamefont
  {Mortimer}}, \bibinfo {author} {\bibfnamefont {Alexander~J}\ \bibnamefont
  {Taylor}}, \bibinfo {author} {\bibfnamefont {Robert}\ \bibnamefont
  {Fickler}}, \bibinfo {author} {\bibfnamefont {Robert~W}\ \bibnamefont
  {Boyd}}, \bibinfo {author} {\bibfnamefont {Mark~R}\ \bibnamefont {Dennis}}, \
  and\ \bibinfo {author} {\bibfnamefont {Ebrahim}\ \bibnamefont {Karimi}},\
  }\bibfield  {title} {\enquote {\bibinfo {title} {Reconstructing the topology
  of optical polarization knots},}\ }\href {\doibase
  https://doi.org/10.1038/s41567-018-0229-2} {\bibfield  {journal} {\bibinfo
  {journal} {Nature Physics}\ }\textbf {\bibinfo {volume} {14}},\ \bibinfo
  {pages} {1079--1082} (\bibinfo {year} {2018})}\BibitemShut {NoStop}%
\bibitem [{\citenamefont {Spreeuw}(1998)}]{spreeuw1998classical}%
  \BibitemOpen
  \bibfield  {author} {\bibinfo {author} {\bibfnamefont {Robert~JC}\
  \bibnamefont {Spreeuw}},\ }\bibfield  {title} {\enquote {\bibinfo {title} {A
  classical analogy of entanglement},}\ }\href {\doibase
  https://doi.org/10.1023/A:1018703709245} {\bibfield  {journal} {\bibinfo
  {journal} {Foundations of physics}\ }\textbf {\bibinfo {volume} {28}},\
  \bibinfo {pages} {361--374} (\bibinfo {year} {1998})}\BibitemShut {NoStop}%
\bibitem [{\citenamefont {Aiello}\ \emph {et~al.}(2015)\citenamefont {Aiello},
  \citenamefont {T{\"o}ppel}, \citenamefont {Marquardt}, \citenamefont
  {Giacobino},\ and\ \citenamefont {Leuchs}}]{aiello2015quantum}%
  \BibitemOpen
  \bibfield  {author} {\bibinfo {author} {\bibfnamefont {Andrea}\ \bibnamefont
  {Aiello}}, \bibinfo {author} {\bibfnamefont {Falk}\ \bibnamefont
  {T{\"o}ppel}}, \bibinfo {author} {\bibfnamefont {Christoph}\ \bibnamefont
  {Marquardt}}, \bibinfo {author} {\bibfnamefont {Elisabeth}\ \bibnamefont
  {Giacobino}}, \ and\ \bibinfo {author} {\bibfnamefont {Gerd}\ \bibnamefont
  {Leuchs}},\ }\bibfield  {title} {\enquote {\bibinfo {title} {Quantum- like
  nonseparable structures in optical beams},}\ }\href {\doibase
  https://doi.org/10.1088/1367-2630/17/4/043024} {\bibfield  {journal}
  {\bibinfo  {journal} {New Journal of Physics}\ }\textbf {\bibinfo {volume}
  {17}},\ \bibinfo {pages} {043024} (\bibinfo {year} {2015})}\BibitemShut
  {NoStop}%
\bibitem [{\citenamefont {Forbes}\ \emph {et~al.}(2019)\citenamefont {Forbes},
  \citenamefont {Aiello},\ and\ \citenamefont
  {Ndagano}}]{forbes2019classically}%
  \BibitemOpen
  \bibfield  {author} {\bibinfo {author} {\bibfnamefont {Andrew}\ \bibnamefont
  {Forbes}}, \bibinfo {author} {\bibfnamefont {Andrea}\ \bibnamefont {Aiello}},
  \ and\ \bibinfo {author} {\bibfnamefont {Bienvenu}\ \bibnamefont {Ndagano}},\
  }\bibfield  {title} {\enquote {\bibinfo {title} {Chapter three - classically
  entangled light},}\ }\href {\doibase
  https://doi.org/10.1016/bs.po.2018.11.001} {\ \bibinfo {series} {Progress in
  Optics},\ \textbf {\bibinfo {volume} {64}},\ \bibinfo {pages} {99 -- 153}
  (\bibinfo {year} {2019})}\BibitemShut {NoStop}%
\bibitem [{\citenamefont {Berg-Johansen}\ \emph {et~al.}(2015)\citenamefont
  {Berg-Johansen}, \citenamefont {T\"{o}ppel}, \citenamefont {Stiller},
  \citenamefont {Banzer}, \citenamefont {Ornigotti}, \citenamefont {Giacobino},
  \citenamefont {Leuchs}, \citenamefont {Aiello},\ and\ \citenamefont
  {Marquardt}}]{Berg-Johansen2015Classically}%
  \BibitemOpen
  \bibfield  {author} {\bibinfo {author} {\bibfnamefont {Stefan}\ \bibnamefont
  {Berg-Johansen}}, \bibinfo {author} {\bibfnamefont {Falk}\ \bibnamefont
  {T\"{o}ppel}}, \bibinfo {author} {\bibfnamefont {Birgit}\ \bibnamefont
  {Stiller}}, \bibinfo {author} {\bibfnamefont {Peter}\ \bibnamefont {Banzer}},
  \bibinfo {author} {\bibfnamefont {Marco}\ \bibnamefont {Ornigotti}}, \bibinfo
  {author} {\bibfnamefont {Elisabeth}\ \bibnamefont {Giacobino}}, \bibinfo
  {author} {\bibfnamefont {Gerd}\ \bibnamefont {Leuchs}}, \bibinfo {author}
  {\bibfnamefont {Andrea}\ \bibnamefont {Aiello}}, \ and\ \bibinfo {author}
  {\bibfnamefont {Christoph}\ \bibnamefont {Marquardt}},\ }\bibfield  {title}
  {\enquote {\bibinfo {title} {Classically entangled optical beams for
  high-speed kinematic sensing},}\ }\href {\doibase 10.1364/OPTICA.2.000864}
  {\bibfield  {journal} {\bibinfo  {journal} {Optica}\ }\textbf {\bibinfo
  {volume} {2}},\ \bibinfo {pages} {864--868} (\bibinfo {year}
  {2015})}\BibitemShut {NoStop}%
\bibitem [{\citenamefont {Hannonen}\ \emph {et~al.}(2019)\citenamefont
  {Hannonen}, \citenamefont {Saastamoinen}, \citenamefont {Lepp{\"a}nen},
  \citenamefont {Koivurova}, \citenamefont {Shevchenko}, \citenamefont
  {Friberg},\ and\ \citenamefont {Set{\"a}l{\"a}}}]{hannonen2019geometric}%
  \BibitemOpen
  \bibfield  {author} {\bibinfo {author} {\bibfnamefont {Antti}\ \bibnamefont
  {Hannonen}}, \bibinfo {author} {\bibfnamefont {Kimmo}\ \bibnamefont
  {Saastamoinen}}, \bibinfo {author} {\bibfnamefont {Lasse-Petteri}\
  \bibnamefont {Lepp{\"a}nen}}, \bibinfo {author} {\bibfnamefont {Matias}\
  \bibnamefont {Koivurova}}, \bibinfo {author} {\bibfnamefont {Andriy}\
  \bibnamefont {Shevchenko}}, \bibinfo {author} {\bibfnamefont {Ari~T}\
  \bibnamefont {Friberg}}, \ and\ \bibinfo {author} {\bibfnamefont {Tero}\
  \bibnamefont {Set{\"a}l{\"a}}},\ }\bibfield  {title} {\enquote {\bibinfo
  {title} {Geometric phase in beating of light waves},}\ }\href {\doibase
  https://doi.org/10.1088/1367-2630/ab3740} {\bibfield  {journal} {\bibinfo
  {journal} {New Journal of Physics}\ }\textbf {\bibinfo {volume} {21}},\
  \bibinfo {pages} {083030} (\bibinfo {year} {2019})}\BibitemShut {NoStop}%
\bibitem [{\citenamefont {Hannonen}\ \emph {et~al.}(2020)\citenamefont
  {Hannonen}, \citenamefont {Partanen}, \citenamefont {Leinonen}, \citenamefont
  {Heikkinen}, \citenamefont {Hakala}, \citenamefont {Friberg},\ and\
  \citenamefont {Set{\"a}l{\"a}}}]{hannonen2020measurement}%
  \BibitemOpen
  \bibfield  {author} {\bibinfo {author} {\bibfnamefont {Antti}\ \bibnamefont
  {Hannonen}}, \bibinfo {author} {\bibfnamefont {Henri}\ \bibnamefont
  {Partanen}}, \bibinfo {author} {\bibfnamefont {Aleksi}\ \bibnamefont
  {Leinonen}}, \bibinfo {author} {\bibfnamefont {Janne}\ \bibnamefont
  {Heikkinen}}, \bibinfo {author} {\bibfnamefont {Tommi~K}\ \bibnamefont
  {Hakala}}, \bibinfo {author} {\bibfnamefont {Ari~T}\ \bibnamefont {Friberg}},
  \ and\ \bibinfo {author} {\bibfnamefont {Tero}\ \bibnamefont
  {Set{\"a}l{\"a}}},\ }\bibfield  {title} {\enquote {\bibinfo {title}
  {Measurement of the pancharatnam--berry phase in two-beam interference},}\
  }\href {\doibase https://doi.org/10.1364/OPTICA.401993} {\bibfield  {journal}
  {\bibinfo  {journal} {Optica}\ }\textbf {\bibinfo {volume} {7}},\ \bibinfo
  {pages} {1435--1439} (\bibinfo {year} {2020})}\BibitemShut {NoStop}%
\bibitem [{\citenamefont {Dudley}\ \emph {et~al.}(2006)\citenamefont {Dudley},
  \citenamefont {Genty},\ and\ \citenamefont
  {Coen}}]{Dudley2006Supercontinuum}%
  \BibitemOpen
  \bibfield  {author} {\bibinfo {author} {\bibfnamefont {John~M.}\ \bibnamefont
  {Dudley}}, \bibinfo {author} {\bibfnamefont {Go\"ery}\ \bibnamefont {Genty}},
  \ and\ \bibinfo {author} {\bibfnamefont {St\'ephane}\ \bibnamefont {Coen}},\
  }\bibfield  {title} {\enquote {\bibinfo {title} {Supercontinuum generation in
  photonic crystal fiber},}\ }\href {\doibase 10.1103/RevModPhys.78.1135}
  {\bibfield  {journal} {\bibinfo  {journal} {Rev. Mod. Phys.}\ }\textbf
  {\bibinfo {volume} {78}},\ \bibinfo {pages} {1135--1184} (\bibinfo {year}
  {2006})}\BibitemShut {NoStop}%
\end{thebibliography}
%

\section*{Acknowledgements}
The authors thank Lauri Salmela for information and data regarding the supercontinuum, and Nikolai V. Tkachenko. 
LK, JDR, MH, TS, MJH, and RF acknowledge the support of the Academy of Finland (Grant No. 308596), the Flagship of Photonics Research and Innovation (PREIN) funded by the Academy of Finland (Grant No. 320165). 
LK, MH and RF acknowledge the support of the Academy of Finland through the Competitive Funding to Strengthen University Research Profiles (Decision 301820).
MH also acknowledges support from the Magnus Ehrnrooth foundation through its graduate student scholarship. 
TS acknowledges Jenny and Arttu Wihuri Foundation for a Ph.D. fellowship.
FB acknowledges support from the National Research Council’s High Throughput Secure Networks challenge program and the Joint Centre for Extreme Photonics.
RF also acknowledges support from the Academy of Finland through the Academy Research Fellowship (Decision 332399). 
\vfill\null

\pagebreak

\section*{Supplementary}
\appendix
\renewcommand{\appendixname}{Supplementary}
\renewcommand\thefigure{S\arabic{figure}}   
\setcounter{figure}{0}  

\section{Tunability}

Several parameters influence the generation of spectral vector beams.
Besides parameters determined by the pulsed Ti:Sapphire laser (such as the spectral bandwidth and the center wavelength), the spectral vector beams can be tuned by changing the polarization basis or by adjusting the orientation and thickness of the birefringent crystal.\\

\begin{figure*}[htb] 
  \centering
  \includegraphics[width= 0.8\textwidth]{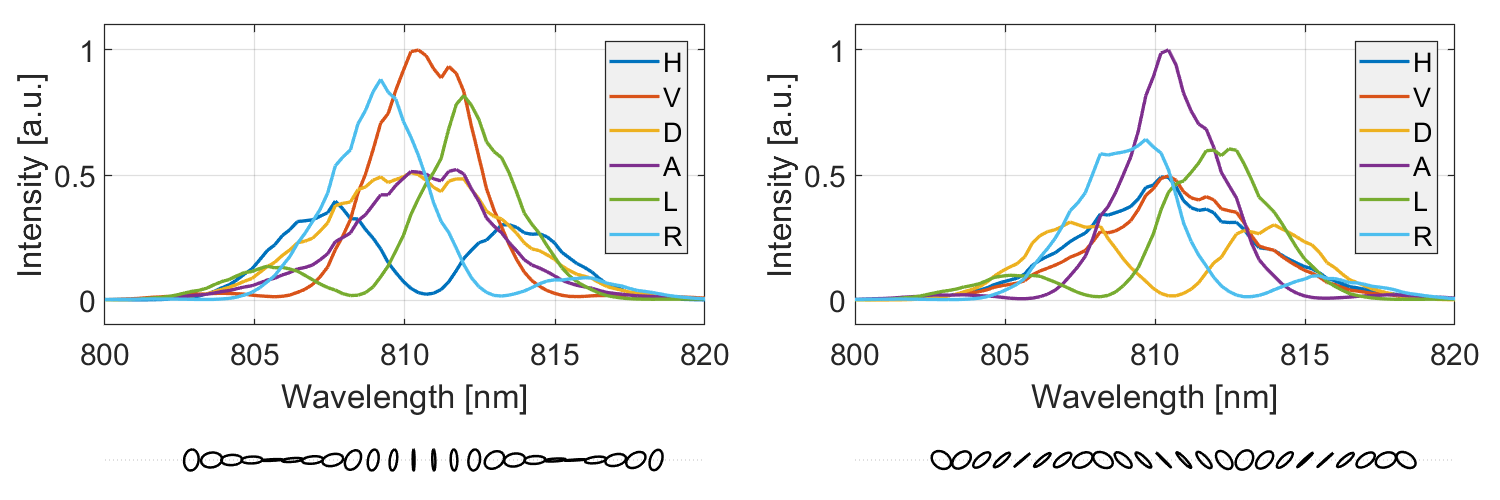}
  \caption{Spectral vector beams with their corresponding polarization patterns, without polarization components in the DA-basis (left) and without polarization components in the HV-basis (right). 
  H, V, D, A, L, R stand for the horizontal, vertical, diagonal, antidiagonal, and left- and right-hand circular polarization components. 
  Note that this notation is used for all figures.}
  \label{fig:bases}
\end{figure*}

In our experimental implementation, the basis in which $\theta$ rotates can be adjusted by choosing a fitting combination and orientation of wave plates.
With a half-wave plate at an angle of 22.5$^\circ$ the spectrum in Fig.~\ref{fig:bases} on the left and without a wave plate the spectrum on the right is generated.
Using a quarter-wave plate at 45$^\circ$ between the fast and slow axis of the wave plate and the axis of polarization of the incident beam the spectrum shown in Fig.~\ref{fig:normal_spectrum} is implemented. 
The latter spectral vector beam only shows linear polarization components simplifying its application in spectroscopic measurements and, thus, was used in all performed experiments.\\

\begin{figure*}[htb] 
  \centering
  \includegraphics[width=0.8\textwidth]{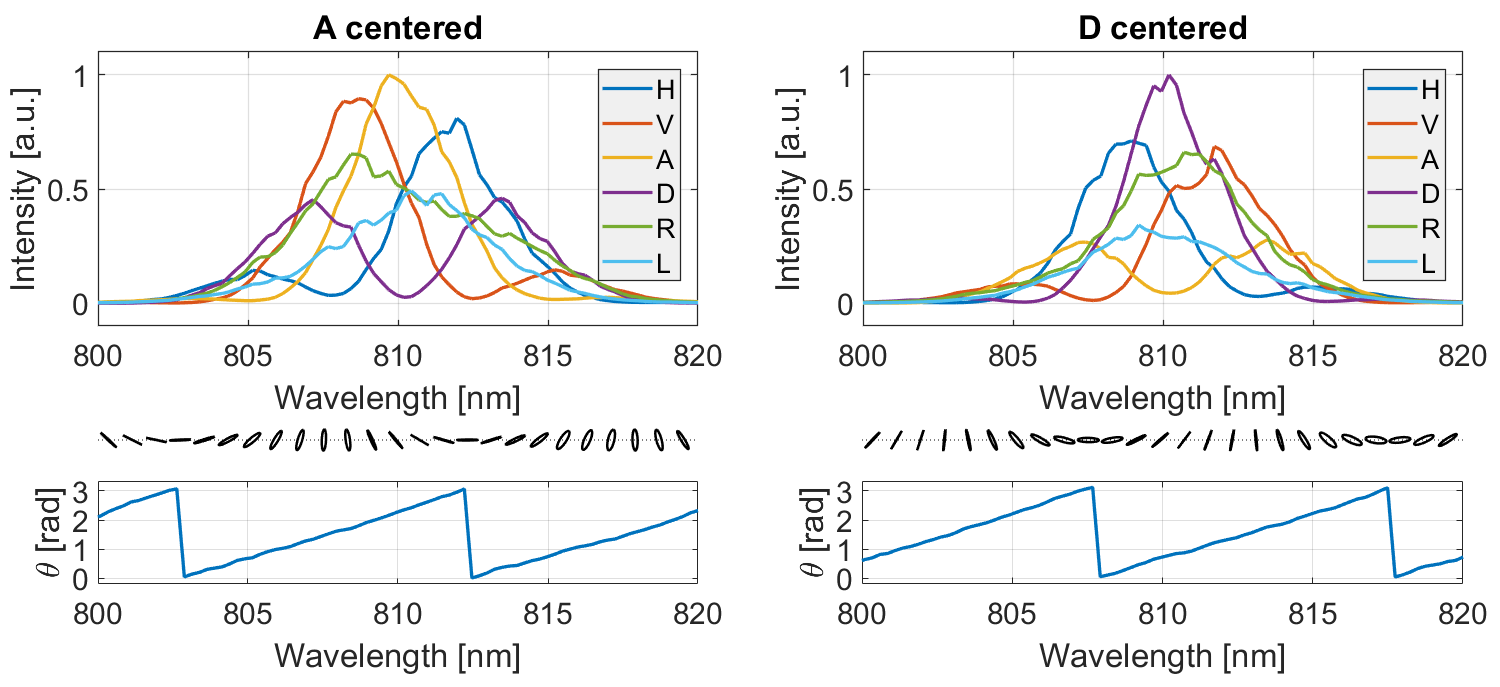}
  \caption{Different crystal orientations influence the polarization patterns. 
  On the left, the crystal is oriented such that the antidiagonal polarization is centered in the spectrum, on the right the diagonal polarization is centered.
  The difference is also visible in $\theta$.}
  \label{fig:crystal_orientation}
\end{figure*}

The impact of different crystal orientations is displayed in Fig.~\ref{fig:crystal_orientation}. 
By rotating the crystal with respect to the vertical axis, thus changing the direction of the optical axis of the birefrigent crystal with respect to the propagation direction of the light field, different polarization components can be centered in the spectrum.\\ 

\begin{figure*}[htb] 
  \centering
  \includegraphics[width=0.9\textwidth]{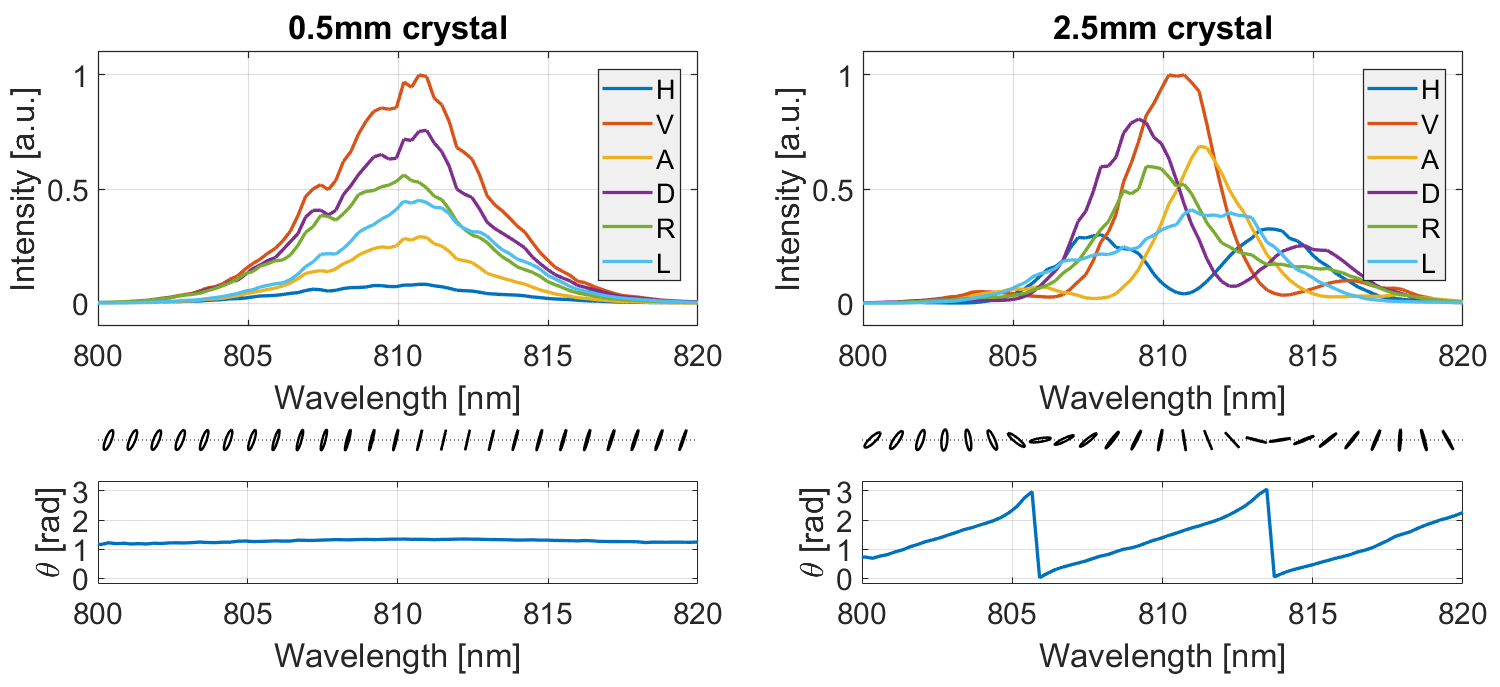}
  \caption{Different crystal thicknesses and how they influence the polarization patterns and rotation angle. 
  }
  \label{fig:crystal_thickness}
\end{figure*}

Finally, the crystal thickness influences the time delay $\tau$ between the two trailing orthogonally polarized components of a pulse and thus the slope of $\theta$.
As shown in Fig.~\ref{fig:crystal_thickness} on the left, a thin crystal generates only a very short time delay and a very shallow slope.
A shallow slope is useful to cover large wavelength bandwidths without risking ambiguity in $\theta$.
Fig.~\ref{fig:crystal_thickness} on the right shows the spectra with a crystal thickness of 2.5\,mm. The gradient of $\theta$ is steep and allows a better resolution for measurements at the cost of a smaller wavelength window without ambiguity.
The spectrum in Fig.~\ref{fig:normal_spectrum} and the spectra used for the measurements in the main text are generated with a 2\,mm thick crystal.
The thickness was chosen to generate a $\pi$ radians rotation (180 degrees) of the polarization ellipse over the predetermined bandwidth of the spectrum without ambiguity.

\section{Degree of polarization}

An additional source of information is the degree of polarization ($p$).
$p$ includes all Stokes parameters
\begin{equation}
p=\frac{\sqrt{S_1^2+S_2^2+S_3^2}}{S_0}
\end{equation}
and is defined from 0 to 1, with 0 being completely unpolarized, 1 completely polarized and everything in between being partially polarized.\\

For a perfectly symmetric spectrum with a pulse width of 200\,fs and a crystal thickness of 2\,mm, $p$ is around 0.14. 
It is nonzero because different polarization components in the beam scale with different intensities due to the temporal shape of the pulse and thus do not cancel out.
By having a very steep gradient of the rotation angle, e.g. many rotations over the spectrum, this can be compensated for.
In this case, $p$ integrated over all wavelengths reaches very low values and the beam is seemingly unpolarized.
Only resolving time or frequency would then reveal the polarization structure.\\

\begin{figure}[htb] 
  \centering
  \includegraphics[width=0.45\textwidth]{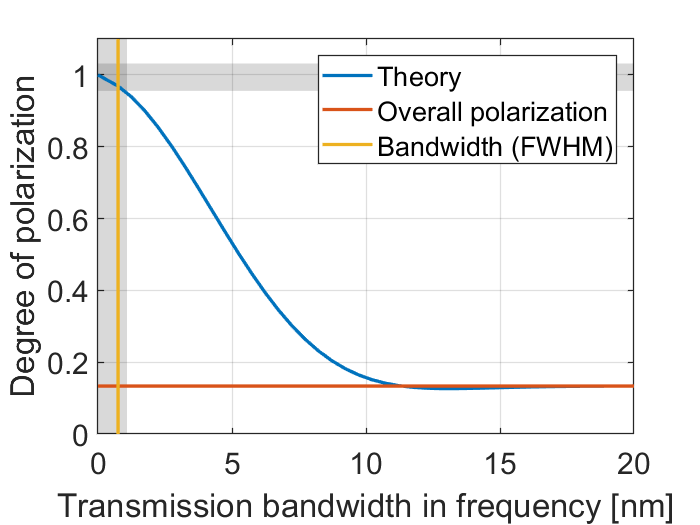}
  \caption{Simulated $p$ for different transmission bandwidths for a filter centered in the spectrum from Fig.~\ref{fig:normal_spectrum}.
  The red line shows the overall $p$ for the spectrum integrated over time and frequency.
  The inaccuracy of $p$ measured with photodiodes (depicted by the grey horizontal bar) translates into an inaccuracy of the transmission bandwidth of 1.1\,nm (depicted by the grey vertical bar).
  The reference bandwidth of 0.76\,nm deduced from measurements with a spectrometer is noted in yellow.}
  \label{fig:DOP}
\end{figure}

The information obtained from evaluating the polarization degree of freedom, for example, can allow to reconstruct the bandwidth of the transmitted light in narrowband transmission measurements .
Fig.~\ref{fig:DOP} shows the simulated curve of $p$ for a transmission band centered in the spectrum from Fig.~\ref{fig:normal_spectrum} as a function of the transmission bandwidth.
However, due to the experimental errors of the value of $p$ measured with our photodiodes, the transmission bandwidth can only be narrowed down to a transmission range of 1.1\,nm, which is roughly 2$\times$ larger than the anticipated transmission bandwidth of 0.76\,nm.
More stable experimental settings could in the future allow exploiting this additional source of information.
For an absorbed frequency band, $p$ is described by a complex function without monotonous behavior and ambiguities, which does not allow the characterization of the absorption bandwidth in a similarly simple way as for the narrowband transmission case.

\section{Evaluating data}

\begin{figure*}[htb] 
  \centering
  \includegraphics[width=0.7\textwidth]{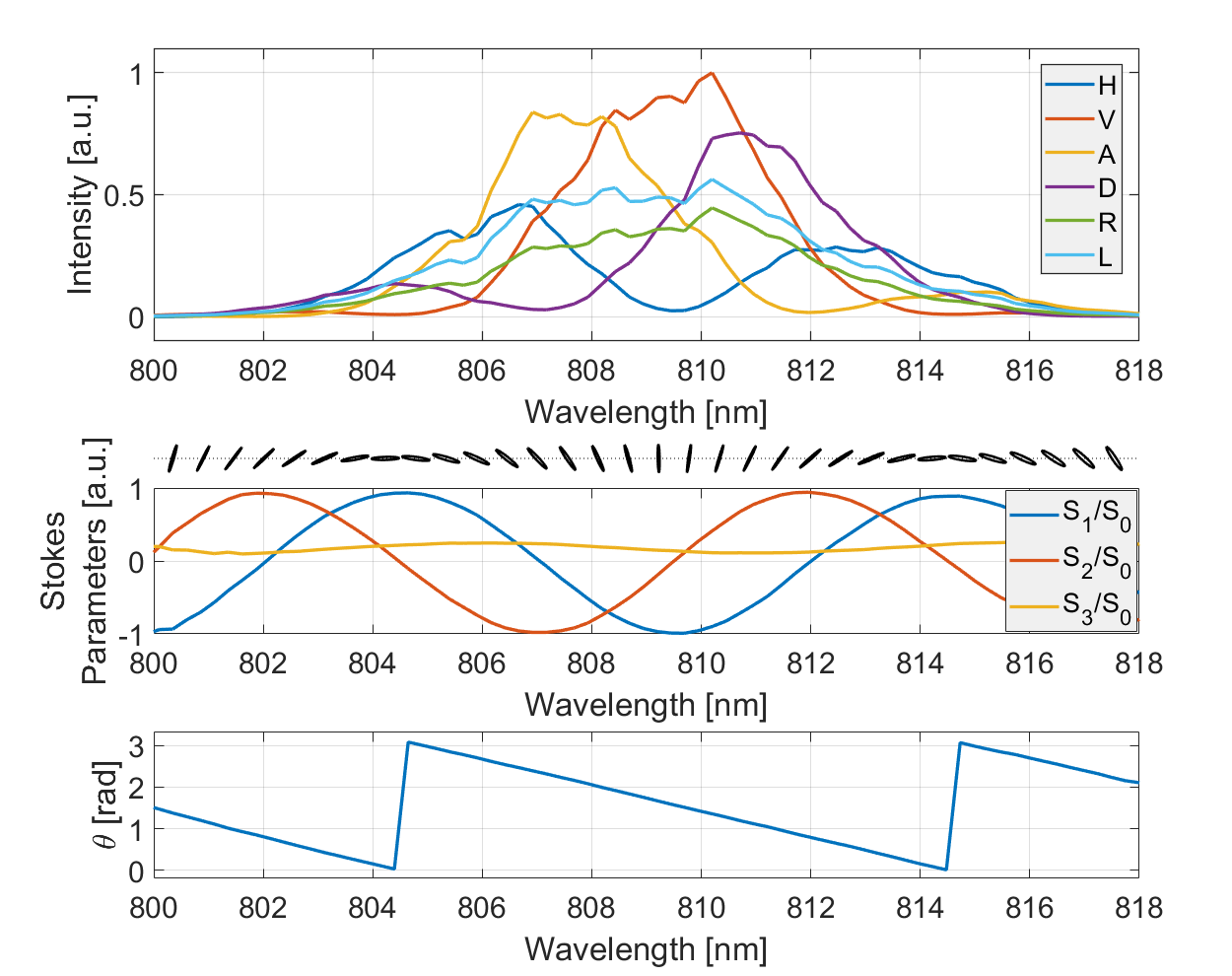}
  \caption{Spectrum of a spectral vector beam with corresponding polarization ellipses, Stokes parameters and rotation angle $\theta$.}
  \label{fig:normal_spectrum}
\end{figure*}

The spectrum from Fig.~2\,b) in the main text is shown in Fig.~\ref{fig:normal_spectrum} with more details. 
The corresponding polarization patterns, the rotation angle $\theta$, and the Stokes parameters are displayed beneath it.\\

The Stokes parameters are defined as
\begin{align}
{S_0}&=\langle E_x^2+E_y^2\rangle = P_{H}+P_{V}\\
{S_1}&=\langle E_x^2-E_y^2\rangle = P_{H}-P_{V}\\
{S_2}&=\langle 2E_x^2 E_y^2\cos \delta\rangle = P_{D}-P_{A}\\
{S_3}&=\langle 2E_x^2 E_y^2\sin \delta\rangle  = P_{R}-P_{L}
\end{align}
with $E$ being the amplitude of the electrical field, the subscripts $x$ and $y$ indicating the standard Cartesian basis, $P$ denoting the power corresponding to the respective polarization in different bases (H=horizontal, V=vertical, D=diagonal, A=antidiagonal, R=right-circular, L=left-circular), and $\delta$ the relative phase between the $x$ and $y$ components.
They are the main measure for the presented characterization of the polarization state.
In the measurements, the spectrometer measures the time-averaged and the photodiodes register the time- and frequency-averaged Stokes parameters per pulse.\\

$\theta$ as introduced in Eq.~3 in the main text is a measure for the ratio of $S_1$ and $S_2$.
It is only defined over a range of $\pi/2$ radians. 
We extend $\theta$ to a range of $\pi$ by taking the spectra's symmetries into account.
If the antidiagonal polarized intensities are larger than the diagonal intensities, a constant factor of $\pi/2$ is added.
This allows covering a larger wavelength range without ambiguity.\\ 

The measurements, in which a center wavelength is reconstructed, follow three main steps.
\begin{enumerate}
\item \textbf{Calibration:} The polarization components of the pulse are measured with a spectrometer to calibrate the system, e.g. as shown in Fig.~\ref{fig:normal_spectrum}.
Then, we match simulations to the experimental data.
These theoretical curves serve as the basis for the evaluation to quantify $\theta(\lambda)$.

\item \textbf{Finding a function for $\theta(\lambda)$:}
For the specific filters such as narrowband transmission or absorption, the expected frequency-integrated polarization angle $\theta$ needs to be calculated.
For the transmission setting, we compute $\theta$ for every wavelength resolved by the spectrometer.
We end up with a function $\theta(\lambda)$ for the frequency and time averaged polarization angle with respect to the filter's transmission wavelength $\lambda$.
$\theta(\lambda)$ of the transmission measurements for example follow the linear function plotted in Fig.~3\,b) in the main text labeled as "Theory".\\
The narrowband absorption is described by a more complex function.
It is found by computing a rectangular block that is absorbing a certain frequency-band with a well-defined width at different positions in the spectrum.
Here, $\theta$ is calculated by integrating over the whole spectrum.
Then, the process at different absorption wavelengths.
The function is displayed in Fig.~3\,e) in the main text.\\
These functions then relate the value of $\theta_{PD}$ measured with the photodiodes to a specific transmission or absorption wavelength of the filter $\lambda$.

\item \textbf{Evaluating the photodiode data:} For different wavelengths, the time- and frequency-averaged $\theta$ is measured with the photodiodes for several pulses.
Its value is compared with the previously defined function $\theta(\lambda)$ to reconstruct the center wavelength of the spectroscopic measurement.
Note, that the absorption measurements have an additional ambiguity (see Fig. 3\,e in the main text). 
To find the right value, we additionally take the $S_0$ value into account.
Above a computed $S_0$ threshold, we choose the wavelength that is closer to the center wavelength of the spectrum, beneath the threshold the wavelength farther away from the center wavelength of the spectrum is chosen.
\end{enumerate}
We note, that once the system is calibrated (step 1) and the function for $\theta(\lambda)$ is found (step 2), only the third step is required for the high-speed spectroscopic measurements presented in the main text.

\section{Supercontinuum}

To show the broadband capabilities, we simulate our method to generate spectral vector beams with an octave-spanning supercontinuum light source.\\
The supercontinuum spanning a wavelength range from 1000 to 2300\,nm ($-40$\,dB bandwidth) is generated by modeling the propagation of sech-type pulses centered at 1555\,nm with 232\,fs pulse duration (FWHM). 
The pulses are injected in the anomalous dispersion regime of a 2\,m nonlinear optical fiber, including third-order dispersion, self-steepening, and Raman effect \cite{Dudley2006Supercontinuum}. 
The nonlinear coefficient of the fiber is $\gamma = 18.4\times 10^{-3}$\,W$^{-1}$m$^{-1}$, and the Taylor-series expansion coefficients for the dispersion at 1550\,nm are $\beta_2= -5.23 \times 10^{-27}$\,s$^2$m$^{-1}$, and $\beta_3 = 4.27\times 10^{-41}$\,s$^3$m$^{-1}$. 
The simulations use 8192 spectral/temporal grid points with a temporal window size of 15\,ps and 10,000 elementary steps along the fiber (step size of 0.2\,mm). 
One-photon-per-mode shot-noise with random phase is added in the frequency domain, although noise effects are not significant in this regime of propagation.\\

\begin{figure*}[htb] 
  \centering
  \includegraphics[width=0.85\textwidth]{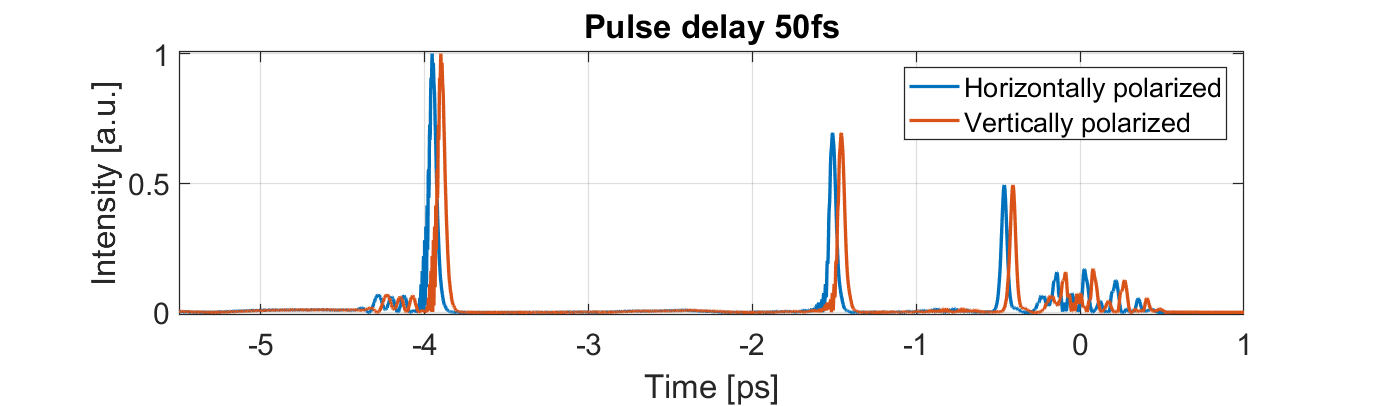}
  \caption{Supercontinuum spectrum in the time domain with a delay of 50\,fs. The undelayed components are horizontally, the delayed components vertically polarized.}
  \label{fig:sc_pulses}
\end{figure*}

\begin{figure*}[htb] 
  \centering
  \includegraphics[width=0.75\textwidth]{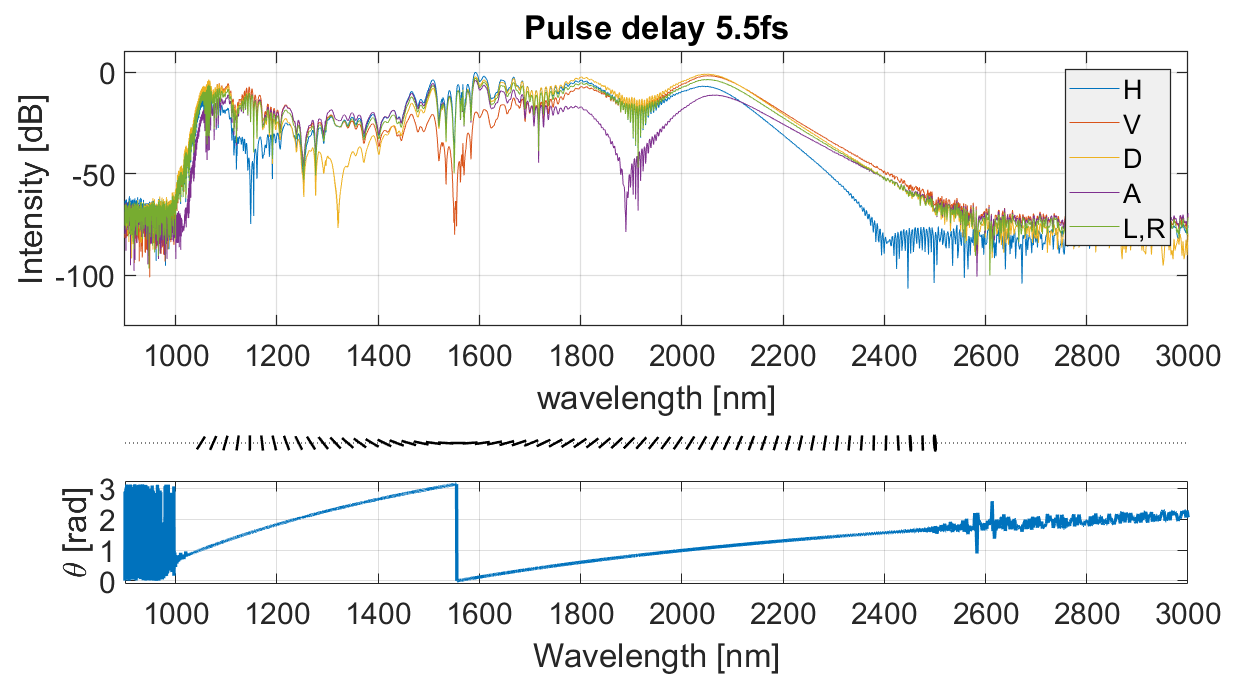}
  \caption{Supercontinuum spectrum in the frequency domain with a 5.5\,fs time delay between the two orthogonally polarized light fields shown in Fig.~\ref{fig:sc_pulses}. The time delay was chosen to cover the supercontinuum bandwidth without ambiguity in $\theta$.}
  \label{fig:sc1}
\end{figure*}

\begin{figure*}[htb] 
  \centering
  \includegraphics[width=0.75\textwidth]{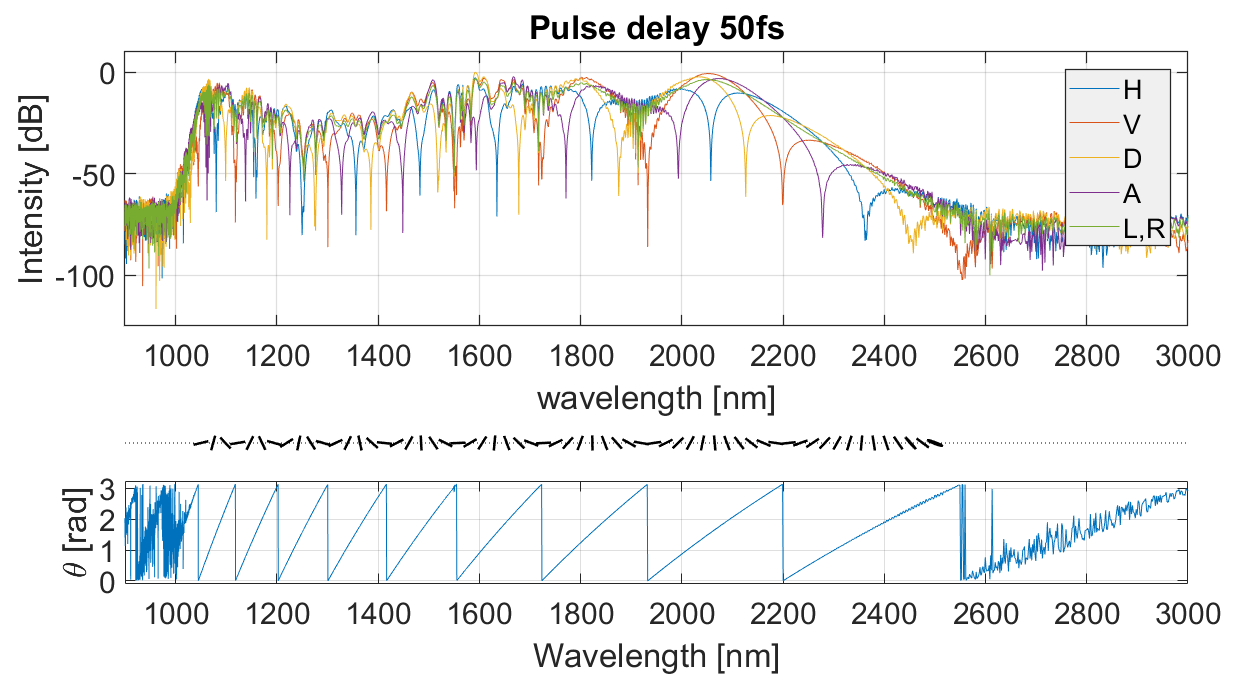}
  \caption{Supercontinuum spectrum in the frequency domain with the  50\,fs time delay displayed in Fig.~\ref{fig:sc_pulses}. This time delay allows a better resolution at the cost of the covered wavelength range or at the cost of ambiguities in $\theta$.}
  \label{fig:sc2}
\end{figure*}

To simulate a supercontinuum spectral vector beam, the complex temporal pulse structure is split up into two parts as shown in Fig.~\ref{fig:sc_pulses}.
One part is horizontally polarized.
The second part is delayed in time and vertically polarized.
They are then superimposed and Fourier transformed to the frequency domain.
As expected, we see a strong correlation between polarization and wavelength as shown in Fig.~\ref{fig:sc1}. 
Similar to the spectral vector beams using pulsed lasers, the delay between the two orthogonally polarized components can be used to control the spectral polarization pattern, e.g. a delay of 50\,fs leads to a linear polarization vector rotating nine times across the full spectrum, as shown in \ref{fig:sc2}. 
The latter pattern can be used to optimize the resolution over the frequency range required in a possible application.  

\end{document}